\documentclass[aps,twocolumn,showpacs,amsmath]{revtex4}
\usepackage{graphicx}
\begin{document}
\setlength{\arraycolsep}{2pt}
\title{Disentanglement of Source and Target and the Laser Quantum State}
\author{Changsuk Noh and H. J. Carmichael}
\affiliation{Department of Physics, University of Auckland, Private Bag 92019,
Auckland, New Zealand} 
\date{\today}

\begin{abstract}
Disentanglement of a laser source from its target qubit is proposed as a
criterion establishing the laser quantum state as a coherent state. It is
shown that the source-target density operator has a {\it unique\/}
factorization in coherent states when the environmental record monitoring
laser pump quanta is ignored. The source-target state conditioned upon the
{\it complete\/} environmental record is entangled, though, as a state of
known total quanta number (source plus target).
\end{abstract}
\maketitle
\narrowtext

We write this letter in response to two recent debates, which, although
they arose and were carried on independently, are shown here to be intimately
connected. The first traces its origin to the observation of Javanainen and Yoo
\cite{Javanainen96} (see also \cite{Cirac96,Chough97}) that the detection of
atoms in overlapping Bose-Einstein condensates produces an interference pattern
``even though the condensates are taken to be in number states with no phases
whatsoever;'' thus, it is not necessary for each condensate to be assigned a
wavefunction of definite phase---specifically, for example, a coherent state.
Subsequently, M\o lmer \cite{Moelmer97} argued for a rather provocative
extrapolation, directed this time at optical rather than material coherence:
``We conjecture that optical coherences, $\ldots$, do not exist $\ldots$.''
and ``A conclusion of this paper is that it does not matter whether coherence
exists or not; observable phenomena in optics and quantum optics are unchanged,
and in this way optical coherences may be regarded as a convenient fiction.''
Elaborated a little, M\o lmer asserts that conventional sources of optical
coherence---lasers---do not (cannot in fact) produce coherent states, though
assigning one is harmless enough---it yields correct answers---and, in
particular, such assignment is ``convenient'' since the more valid (nonfictional)
calculation requires one to work with entangled states.

Unsurprisingly, not all readers of M\o lmer agreed \cite{Gea-Banacloche97,Moelmer98}.
In retrospect, however, his and Javanainen and Yoo's comments are not as new
or radical as it seems. If, to simplify, we focus the issue on whether number
states or coherent states give the more fundamental account of laser light,
then the question is raised already by the earliest laser theories: the
Scully-Lamb master equation \cite{Scully66} describes a birth-death evolution
between number states---$\ldots|n-1\rangle\to|n\rangle\to|n+1\rangle \to|n\rangle\ldots$
as a stochastic process---while the phase-space formulations of the Lax
and Haken schools \cite{Lax67b,Graham70} underlie the popular picture of
a stochastically evolving coherent state, $|\alpha_t\rangle$, with  $\alpha_t$
a Brownian path mapped out in the complex plane. 

In a final chapter, this first debate was carried into the quantum information
field by Rudolph and Sanders \cite{Rudolph01}, who criticize the claimed
implementation of continuous variable quantum teleportation (CVQT) by Furusawa
{\it et al.} \cite{Furusawa98}. They argue that ``genuine CVQT cannot be
achieved using conventional laser sources, due to an absence of intrinsic
coherence'' and for this cite the assertion of M\o lmer \cite{Moelmer97}
(see \cite{vanEnk01} for a response).

The second, related debate arose within the quantum information field itself,
initiated by Geo-Banacloche \cite{Gea-Banacloche02} and van Enk and Kimble
\cite{vanEnk02} who point out that the state of a laser manipulating a qubit
is (or potentially is) entangled, in which case decoherence arising from the
resolution of the entanglement (tracing over the laser) must be considered
when assessing the fidelity of the manipulation. M\o lmer, in the quoted paper,
 makes a similar point: ``the state of an atomic target for a laser beam is
entangled with the field states, as inferred from  the Hamiltonian (1).''
From the other side, Itano \cite{Itano03} objects, arguing, principally,
following Mollow \cite{Mollow75}, that the Hamiltonian of a two-state system
interacting with a quantum field in a coherent state---say $|\alpha_t\rangle$---may
be taken with operators replaced by classical parameters, e.g., the annihilation
operator by $\alpha_t$. He asserts, in particular, that no source of decoherence
is missed when this is done; {\it all\/} decoherence of the qubit arises through
spontaneous emission (see \cite{Gea-Banacloche03} and \cite{vanEnk03} for replies
to Itano's comment).

This letter is most directly an extension of the contribution of one of us to
the second debate. Taking a different approach to Itano, Nha and Carmichael
\cite{Nha05} use the quantum trajectory theory of cascaded open systems
\cite{Carmichael93b} to reach his conclusion;
assuming the laser produces a coherent state in the absence of the target
qubit---say $|\alpha_t\rangle$---they show that a factorized state is reached
in its presence, a state $|\alpha_t\rangle\otimes|T(t)\rangle$, with $|T(t)\rangle$
the target state. Here, building upon this result, we bring the two debates together.
First, Nha and Carmichael's {\it assumption\/} of a coherent state is relaxed;
the laser is modeled realistically in the spirit of \cite{Scully66,Lax67b,Graham70},
though the work of Rice and Carmichael \cite{Rice94} outlines, more simply, the
essential features of the laser model. We begin then with the question: does a
{\it realistic\/} laser source entangle with its target qubit? We argue, yes,
but only so long as the source-target state is conditioned upon the entire
environmental record, including the record of the laser pumping (we adopt the
notion of entangled conditional states \cite{Nha04}).  Discarding the pumping
record, so the source-target state is mixed, the conclusion does not follow.
Rather we show that the resulting mixed state is factorizable, and, if the source
is classical (possesses a nonsingular and positive Glauber-Sudarshan $P$ function),
the factorization is {\it unique} and assigns the state of the laser source as a
coherent state.

Before we elaborate these results a comment is in order regarding the nature
of the problem we aim to address and our strategy for an answer. Griffiths \cite{Griffiths84}
states the problem as well as it can be stated: ``In essence, nonrelativistic
quantum mechanics consists in solving the Schr\"odinger equation and giving a
physical interpretation to the solutions (including boundary and initial conditions).
The former is a mathematical problem about which there is little disagreement.
The latter has given rise to an extended controversy which is far from being
resolved.'' We might add that much of the controversy stems from a mixing of the
former with the latter, and emphasize that interpretation must, indeed, be
{\it given}; the Schr\"odinger equation will not serve up the state of laser
light on its own: even on a cosmic scale it merely entangles---and
nothing ever happens (it says nothing of events). We quote from Griffiths'
paper on consistent histories (see also Griffiths \cite{Griffiths93}, Omn\` es
\cite{Omnes88}, and Gell-Mann and Hartle \cite{Gell-Mann90}). Without directly
building upon that program, we understand the problem of interpretation much
as it does, see histories---trajectories (sequences of events or assignable
properties)---as the key element of an answer, and, most importantly for this
letter, place {\it consistency\/} at the center of our argument.

The apparent consensus in quantum optics is that the question of the state of
laser light ``ends up being of only academic interest'' \cite{Moelmer98} and
``which way one chooses to resolve a particular density operator---i.e., which
states one chooses to ascribe to individual realizations of the ensemble---may
ultimately be little more than a matter of convenience'' \cite{Gea-Banacloche97};
to claim otherwise is a commission of the ``partition ensemble fallacy'' \cite{Rudolph01}.
Our claim in this letter {\it is} otherwise. We place the question in the
category of those addressed by decoherent histories, not merely academic,
but of deep importance to a coherent formulation of quantum mechanics. We
claim, moreover, that criteria exist to assign the laser state uniquely as
a coherent state. The main points of our argument, elaborated below, are:
(i) the output of a light source is intended for illuminating a target, therefore
the principal criterion in assigning a state is to disentangle the source and
target so the behavior of the latter may be seen to follow from properties thus
assigned the former; (ii) such disentanglement is not always feasible but
is for any classical photoemissive source if the description is suitably
coarse-grained---i.e., if information in the environment allowing for the
tracking of source-target correlations at the one-quantum level is discarded;
(iii) coherent states provide the only self-consistent disentanglement;
by this we mean that with the factorization adopted, a detailed description
of the target response can be given, as a stochastic process, preserving the
factorization at every step.

While (i) is central, (iii) is also essential to escape the partition ensemble
fallacy, which, from \cite{Rudolph01}, states: ``Within the framework of standard
quantum mechanics, the density operator is the complete description of the
quantum state [10], and there is no reason to accord preferential treatment
to one particular decomposition of the infinite number of equivalent
decompositions for a mixed state.'' Against this ``behavior'' in (i)
and ``stochastic process'' in (iii) point to the fundamental oversight of the
quoted ``fallacy'': the density operator {\it does not\/} provide a complete
description of laser light as a quantum field; an infinity of correlation
functions do that. We assert that coherent states disentangle the laser source
and target not only in their state $\rho(t)$, but as a stochastic process,
correlation functions to all orders taken into account.

\begin{figure}[b]
\includegraphics*[width=3in,keepaspectratio=true]{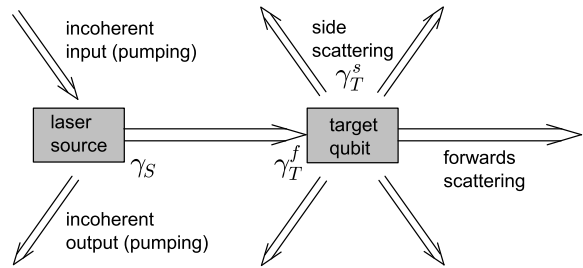}
\caption{Schematic of the scattering scenario for a laser source illuminating
a target qubit.}
\label{fig:fig1}
\end{figure}

We build our argument around the model scattering scenario depicted in
Fig.~\ref{fig:fig1} which we describe by the  master equation in Lindblad form
\cite{Carmichael93b,Kolobov87,Gardiner93}
\begin{equation}
\frac{d\rho}{dt}=({\mathcal L}_{S}+{\mathcal L}_T+{\mathcal L}_{ST})\rho,
\label{eqn:master_equation1}
\end{equation}
where ${\mathcal L}_S$ depends upon the particular laser model used and
(setting $\hbar=1$)
\begin{eqnarray}
{\mathcal L}_T&=&-i[\hat H_T,\cdot\,]+{\textstyle\frac12}
(2\hat s\cdot\hat s^\dagger-\hat s^\dagger\hat s\cdot-\cdot\hat s^\dagger\hat s),\\
{\mathcal L}_{ST}&=&-i[\hat H_{ST},\cdot\,]+{\textstyle\frac12}
(2\hat f\cdot\hat f^\dagger-\hat f^\dagger\hat f\cdot-\cdot\hat f^\dagger\hat f),
\end{eqnarray}
with coupling Hamiltonian
\begin{equation}
\hat H_{ST}=i{\textstyle\frac12}\sqrt{\gamma_S\gamma_T^f}(\hat a^\dagger\hat b
-\hat a\hat b^\dagger),
\label{eqn:coupling_Hamiltonian}
\end{equation}
and jump operators
\begin{equation}
\hat f=\sqrt{\gamma_S}\hat a+\sqrt{\gamma_T^f}\hat b,\qquad
\hat s=\sqrt{\gamma_T^s}\hat b;
\label{eqn:jump_operators}
\end{equation}
operator $\hat a$ annihilates photons from the laser mode and $\hat b$ lowers the
target qubit. The strength of the one-way coupling is specified through bandwidths,
$\gamma_S$ and $\gamma_T^f$, of the source output channel and target input channel,
and the target scatters light with Einstein $A$ coefficient $\gamma_T=\gamma_T^f+\gamma_T^s$
(forwards plus side); see van Enk \cite{vanEnk04} for the generalization of this
simplified two-channel model.

The one-way nature of the coupling (scattering character) is made explicit in a quantum
trajectory unraveling of Eq.~(\ref{eqn:master_equation1}) \cite{Carmichael93b}, which
replaces $\hat H_{ST}$ by the non-Hermitian Hamiltonian
\begin{eqnarray}
\hat H_{ST}-i{\textstyle\frac12}\hat s^\dagger\hat s-i{\textstyle\frac12}
\hat f^\dagger\hat f&=&-i\sqrt{\gamma_S\gamma_T^f}\hat a\hat b^\dagger\nonumber\\
&&-i{\textstyle\frac12}\gamma_S\hat a^\dagger\hat a-i{\textstyle\frac12}\gamma_T
\hat b^\dagger\hat b.
\label{eqn:non-Hermitian_Hamiltonian}
\end{eqnarray}
It is essential as it then follows from Eqs.~(\ref{eqn:jump_operators}) and
(\ref{eqn:non-Hermitian_Hamiltonian}), assuming a coherent state, $|\alpha_t\rangle$,
for the source, that the conditional state {\it factorizes\/} as \cite{Nha05}
\begin{equation}
|\psi_{\rm REC}(t)\rangle=|\alpha_t\rangle\otimes|T(t)\rangle,
\label{eqn:conditional_state}
\end{equation}
with $|T(t)\rangle$ the target state. The result requires the cancelation of
$\hat a^\dagger\hat b$ from $\hat H_{ST}$ and the special status of coherent
states as eigenstates of $\hat a$.

Now let us remove the  coherent state assumption and allow the laser to be
described, dynamically, through  ${\cal L}_S$. Generally, details of what follows
depend on the laser model. We specify it here in the simplest form to accommodate
our aims---i.e., as a birth-death process for carrier (gain medium) and photon (laser
light) numbers $N_t$ and $n_t$ \cite{Rice94}. A central feature is the {\it incoherent\/}
pumping. At this input-output channel (Fig.~\ref{fig:fig1}) the passage of energy
quanta into and out of the laser resonator may be tracked and corresponding changes
in $N_t$ inferred. As a simplifying feature, we assume the gain-medium polarization
may be adiabatically eliminated so  $n_t$ changes according to a birth-death process
too: e.g., under stimulated emission, $(N_t,n_t)\to(N_t-1,n_t+1)$. If, then, the side
and forwards scattering of the target qubit is monitored, thus completing the scattering
record, the conditional state is an {\it entangled\/} state of photon number and qubit
excitation,
\begin{equation}
|N_t\rangle\otimes\left[a(t)|n_t\rangle\otimes|-\rangle+
b(t)|n_t-1\rangle\otimes|+\rangle\right],
\label{eqn:entangled_state}
\end{equation}
where $a(t)$ and $b(t)$ are real coefficients.

At this point our rudimentary model might be criticized as, for example, it rules
out entanglement of the laser medium and its emitted light. Recall, however, that
breaking the unending chain of entanglement is the goal, and something recognized
as necessary---if named a ``convenient fiction''---also by M\o lmer \cite{Moelmer97}.
Recall that interpretation must be {\it given}. We argue in this spirit that
coherent states are special, indeed unique: they assign properties, separately
and self-consistently, to the laser source and target qubit.

We now discard the pumping record that permits the labeling of conditional state
(\ref{eqn:entangled_state}) by definite numbers, $N_t$ and $n_t$; although, in
principle, the tracking of every quantum in the environment might be permitted,
it is certainly impossible to achieve in practice. Thus, we now consider a
coarse-grained description, summing over possible carrier numbers to arrive at
a source-target density operator of the form
\begin{equation}
\rho(t)=\sum_{N=0}^\infty p_N(t)\rho^{(N)}(t),
\label{eqn:density_matrix}
\end{equation}
where $\rho^{(N)}$ has non-zero matrix elements [Eq.~(\ref{eqn:entangled_state})] 
\begin{equation}
\rho^{(N)}_{n,-;n,-},\quad\rho^{(N)}_{n-1,+;n-1,+},\quad\rho^{(N)}_{n,-;n-1,+},
\quad\rho^{(N)}_{n-1,+;n,-}.
\label{eqn:non-zero_elements}
\end{equation}
The question follows: do these equations specify an entangled or a factorizable
state? The answer by construction is a factorizable state, since the same set of
nonzero elements is recovered by the expansion
\begin{equation}
\rho^{(N)}=\frac1{2\pi}\int P(r)|re^{i\phi}\rangle\langle re^{i\phi}|\otimes
|T_\phi\rangle\langle T_\phi|rdrd\phi,
\label{eqn:factorized_state} 
\end{equation}
where $|re^{i\phi}\rangle$ is a coherent state of amplitude $r$ and phase
$\phi$, $P(r)$ is a distribution over $r$, and
\begin{equation}
|T_\phi\rangle=a^\prime e^{-i\phi/2}|-\rangle+b^\prime e^{i\phi/2}|+\rangle
\end{equation}
is a target state. The phase average is the crucial thing, and it suggests that
the principal function of the entangled state (\ref{eqn:entangled_state}) is to
account for a phase correlation between source and target when the known numbers
$N_t$ and $n_t$ disallow a separable phase assignment. If, on the other hand,
the pumping record is discarded, the correlation between $n_t$ and qubit
excitation is discarded too. Only the phase correlation remains---the principal
signature of the macroscopic physics. This, coherent states can capture in an
assigned property of the source---complex amplitude $re^{i\phi}$---and a
correlated target response. Such an interpretation realizing the separation
of source and target is clearly desired, if not required, since (\ref{eqn:density_matrix})
is indeed a factorizable state. It remains to show its generality and its uniqueness.

With Eq.~(\ref{eqn:factorized_state}) as our motivation, we aim to show that the
solution, $\rho(t)$, to master equation (\ref{eqn:master_equation1}) factorizes
in coherent states whenever the source is classical---i.e., whenever its density
operator possesses a nonsingular and positive Glauber-Sudarshan $P$ representation.
To this end, we regroup terms in the superoperator ${\mathcal L}_{S}+{\mathcal L}_T
+{\mathcal L}_{ST}$ reexpressing it as
${\mathcal L}_{S}^\prime+{\mathcal L}_T^\prime+{\mathcal L}_{ST}^\prime$,
where ${\mathcal L}_S^\prime$ and ${\mathcal L}_T^\prime$ operate on the source
or target only,
\begin{eqnarray}
{\mathcal L}_S^\prime&=&{\mathcal L}_S+{\textstyle\frac12\gamma_S(2\hat a\cdot
\hat a^\dagger-\hat a^\dagger\hat a\cdot-\cdot\hat a^\dagger\hat a)},\\
{\mathcal L}_T^\prime&=&-i[\hat H_T,\cdot\,]+{\textstyle\frac12}\gamma_T
(2\hat b\cdot\hat b^\dagger-\hat b^\dagger\hat b\cdot-\cdot\hat b^\dagger\hat b),
\end{eqnarray}
and all cross terms are collected in the coupling
\begin{eqnarray}
{\mathcal L}_{ST}^\prime&=&-i[\hat H_{ST},\cdot\,]+{\textstyle\frac12}\sqrt{\gamma_S
\gamma_T^f}(2\hat a\cdot\hat b^\dagger+2\hat b\cdot\hat a^\dagger\nonumber\\
&&-\hat a^\dagger\hat b\cdot-\hat b^\dagger\hat a\cdot-\cdot\hat a^\dagger\hat b
-\cdot\,\hat b^\dagger\hat a)\nonumber\\
&=&\sqrt{\gamma_S\gamma_T^f}(\hat a\cdot\hat b^\dagger+\hat b\cdot\hat a^\dagger
-\hat b^\dagger\hat a\cdot-\cdot\hat a^\dagger\hat b).
\label{eqn:irreversible_coupling}
\end{eqnarray}
We also need the master equation traced over the qubit,
\begin{equation}
\frac{d\rho_S}{dt}={\mathcal L}_S^\prime\rho_S,\qquad\rho_S={\rm tr}_T(\rho),
\label{eqn:master_equation3} 
\end{equation}
assumed to possess the solution for a classical source
\begin{equation}
\rho_S(t)=\int P(\alpha,\alpha^*,t)|\alpha\rangle\langle\alpha|d^2\alpha,
\label{eqn:diagonal_representation}
\end{equation}
where $P(\alpha,\alpha^*,t)$ is a nonsingular and positive function; most likely
the solution is coarse-grained---e.g., through a system size expansion---but
it is acceptable in the spirit of the discarded pumping record. For simplicity,
all dependence on the gain medium---e.g., $|N_t\rangle$ in (\ref{eqn:entangled_state})---is
omitted. Note now that $P(\alpha,\alpha^*,t)$ may be written as a
functional integral,
\begin{equation}
P(\alpha,\alpha^*,t)=\int\delta^{(2)}(\alpha-\alpha_t)P^\prime(\alpha_t)d^2\alpha_t,
\label{eqn:integral_over_paths}
\end{equation}
\begin{subequations}
with $P^\prime(\alpha_t)$ a distribution over stochastic paths $\alpha_t$. Thus,
as a generalization of Eq.~(\ref{eqn:factorized_state}), we propose the
{\it ansatz\/}
\begin{eqnarray}
\rho(t)&=&\int P(\alpha,\alpha^*,t)|\alpha\rangle\langle\alpha|\otimes\rho_{T|\alpha}
(t)d^2\alpha\label{eqn:ansatza}\\
&=&\int P^\prime(\alpha_t)|\alpha_t\rangle\langle\alpha_t|\otimes\rho_{T|\alpha_t}
(t)d^2\alpha_t,
\label{eqn:ansatzb}
\end{eqnarray}
where $\rho_{T|\alpha_t}(t)$ is the target state conditioned upon $\alpha_t$.
\end{subequations}

It follows readily that the {\it ansatz\/} satisfies the master equation.
The result follows from the one-way coupling (scattering character)
as in our comment below Eq.~(\ref{eqn:conditional_state}): ``The result requires the
cancelation of $\hat a^\dagger\hat b$ from $\hat H_{ST}$ and the special
status of coherent states as eigenstates of $\hat a$.'' Here the cancelation yields
a propitious ordering of operators in Eq.~(\ref{eqn:irreversible_coupling})---every
$\hat a$ acting from the left and $\hat a^\dagger$ from the right. With this
and (\ref{eqn:ansatzb}), the  source-target coupling is 
\begin{equation}
{\mathcal L}^\prime_{ST}\{|\alpha_t\rangle\langle\alpha_t|\otimes\rho_{T|\alpha_t}
(t)\}=-i[\hat H_{\rm drive}(\alpha_t),\rho_{T|\alpha_t}(t)],
\label{eqn:relation1}
\end{equation}
with
\begin{equation}
\hat H_{\rm drive}(\alpha_t)=i\sqrt{\gamma_S\gamma_T^f}(\alpha_t^*\hat b
-\alpha_t\hat b^\dagger),
\end{equation}
where the quantized one-way interaction is replaced by a symmetric
interaction of the target qubit with a classical field (recall Mollow \cite{Mollow75}).
The proof is completed by the further result, using Eqs.~(\ref{eqn:master_equation3})
and (\ref{eqn:diagonal_representation}),
\begin{equation}
\int\left[\frac{\partial P(\alpha,\alpha^*,t)}{\partial t}|\alpha\rangle\langle\alpha|
-P(\alpha,\alpha^*,t)({\mathcal L}^\prime_S|\alpha\rangle\langle\alpha|)\right]d^2\alpha=0.
\label{eqn:relation2}
\end{equation}
Substituting $\rho$ in the form (\ref{eqn:ansatza}) [(\ref{eqn:ansatzb})] in $d\rho/dt$ and
${\mathcal L}_S^\prime\rho$ [${\mathcal L}_T^\prime\rho$ and ${\mathcal L}_{ST}^\prime\rho$],
we 
find the 
master equation solved if $\rho_{T|\alpha_t}$
satisfies the separated target equation 
\begin{eqnarray}
\frac{d\rho_{T|\alpha_t}}{dt}&=&{\mathcal L}_T^\prime\rho_{T|\alpha_t}-i[\hat H_{\rm drive}
(\alpha_t),\rho_{T|\alpha_t}].
\label{eqn:target_master_equation}
\end{eqnarray}

It is important to state in conclusion that Eqs.~(19a,b) and (\ref{eqn:target_master_equation})
separate not only a state, $\rho(t)$, but a stochastic process, fully characterizing the
dynamics of the source as a quantum field and the target response. At the level of the
master equation, this follows from the Markov assumption, which permits the extension
from $\rho(t)$ to correlation functions (Lax's quantum regression \cite{Lax63}). It is
particularly clear in quantum trajectories, where, returning to Eqs.~(\ref{eqn:jump_operators})
and (\ref{eqn:non-Hermitian_Hamiltonian}), the coherent state factorization is
self-consistently preserved if, once adopted, the target qubit quantum jumps are
tracked. In this, as eigenstates of $\hat a$, coherent states are unique.

Even in an extended system the Schr\"odinger equation entangles every system part with
every other. If, however, one gives up an account at the level of every single quantum
the resulting mixed state might factorize. We, thus, separated the state of a laser
source and its target. We propose this separability as a criterion to assign
the laser state uniquely as a coherent state.

The authors acknowledge discussions with Hyunchul Nha
and support from the Marsden Fund of the RSNZ.


\end{document}